\documentstyle[12pt,fleqn]{article}
\input epsf
\def\singlespace {\smallskipamount=3.75pt plus1pt minus1pt
                  \medskipamount=7.5pt plus2pt minus2pt
                  \bigskipamount=15pt plus4pt minus4pt
                  \normalbaselineskip=15pt plus0pt minus0pt
                  \normallineskip=1pt
                  \normallineskiplimit=0pt
                  \jot=3.75pt 
                  {\def\smallskip {\vskip\smallskipamount}}
                  {\def\medskip   {\vskip\medskipamount}}
                  {\def\bigskip   {\vskip\bigskipamount}}
                  {\setbox\strutbox=\hbox{\vrule 
                    height10.5pt depth4.5pt width 0pt}}
                  \parskip 7.5pt
                  \normalbaselines}

\def\doublespace {\smallskipamount=7.5pt plus2pt minus2pt
                  \medskipamount=15pt plus4pt minus4pt
                  \bigskipamount=30pt plus8pt minus8pt
                  \normalbaselineskip=30pt plus0pt minus0pt
                  \normallineskip=2pt
                  \normallineskiplimit=0pt
                  \jot=7.5pt
                  {\def\smallskip {\vskip\smallskipamount}}
                  {\def\medskip   {\vskip\medskipamount}}
                  {\def\bigskip   {\vskip\bigskipamount}}
                  {\setbox\strutbox=\hbox{\vrule 
                    height21.0pt depth9.0pt width 0pt}}
                  \parskip 15.0pt
                  \normalbaselines}

\include{dspace12}
\def\be{\begin{equation}}
\def\ee{\end{equation}}
\def\bea{\begin{eqnarray}}
\def\eea{\end{eqnarray}}

\def\sect #1{\setcounter{equation}{0}}

\begin{document}
\singlespace

\begin{center}
{\Large {On the global visibility of singularity in quasi-spherical collapse}}
\end{center}
\vspace{1.0in}
\vspace{12pt}

{\begin{center}
{\large{ S. S. Deshingkar\footnote{E-mail : shrir@tifrvax.tifr.res.in}, S.
Jhingan\footnote{E-mail : sanju@tifrvax.tifr.res.in} and P. S. 
Joshi\footnote{E-mail : psj@tifrvax.tifr.res.in}\\
Theoretical Astrophysics Group\\
Tata Institute of Fundamental Research\\
Homi Bhabha Road, Colaba, Mumbai 400 005, India.}} 
\end{center}

\vspace{1.3in}

\begin{abstract}

We analyze here the issue of local versus the global visibility of a
singularity that forms in gravitational collapse of a dust cloud, which has
important implications for the weak and strong versions of the cosmic
censorship hypothesis. We find conditions as to when a singularity will be
only locally naked, rather than being globally visible, thus preseving the 
weak censorship hypothesis. The conditions for formation of a black hole or 
naked singularity in the Szekeres quasi-spherical collapse models are worked
out. The causal behaviour of the singularity curve is studied by examining 
the outgoing radial null geodesics, and the final outcome of collapse is 
related to the nature of the regular initial data specified on an initial 
hypersurface from which the collapse evolves. An interesting feature that 
emerges is the singularity in Szekeres spacetimes can be ``directionally 
naked''.

\end{abstract}

\newpage
\doublespace
\section{Introduction} 

An important question regarding the final configurations resulting out
of gravitational collapse of massive matter clouds is, in case a naked
singularity develops rather than a black hole, whether it would be visible to
a far away observer or not. This issue of local versus global visibility
of the singularity has important implications for the cosmic censorship 
principle. Whenever naked singularities formed, if they were only locally
naked rather than being globally visible, that would mean the weak cosmic 
censorship would hold, even though the strong censorship may be violated
\cite{Pen}.  In its strong version the cosmic censorship states that 
singularities cannot be even locally visible, whereas the weak formulation
allows for the singularities to be at least locally naked, but requires that 
they should not be globally naked. 

If the naked singularities formed but were always only locally naked and 
not visible to far away observers outside the event horizon, then though the 
strong form is violated, the validity of the weak version has 
the astrophysical implication that far away observers may never see the 
emissions from the singularity, or the very dense strong curvature regions 
near the same. This effectively seals the singularity inside the black hole. 
Thus, given a gravitational collapse scenario, it is of importance 
to determine 
if the singularity forming could be only locally naked, and if so under what 
conditions. On the other hand, if the singularity is globally visible, thus 
accessible to the outside observers, one would like to ascertain the conditions
under which this would be possible, and to relate these to the nature of 
regular initial data from which the collapse develops. 

Our purpose here is to analyze a wide class of collapse models, namely
the quasi-spherical spacetimes given by Szekeres\cite{Szeke}, from such a 
perspective. Such an investigation is also of additional 
interest because these models 
are not spherically symmetric. The non-solvability of Einstein equations 
restrict most of the studies of formation and structure of singularities in 
gravitational collapse limited to spherically symmetric models, and a simple 
enough equation of state, namely $p=0$. 
Though results on some specific solutions are available 
for more general equation of state $p=k\rho$\cite{Jhingan}, still most of 
the work considering departures from spherical symmetry is numerical (see
e.g. \cite{Shapiro}). 
Analytic study of collapsing models with more general geometries 
is of considerable interest since strict spherical symmetry is a strong 
assumption. Thus we study here the formation of singularities in the 
Szekeres spacetimes, which do not admit any Killing vectors but allow an 
invariant family of two spheres, and has the 
Tolman-Bondi-Lema{\^i}tre (TBL) model
\cite{Tolman} as its special case. In a recent paper Joshi and Kr{\'o}lak
\cite{JosKro} have shown existence of naked singularities in these models. 
Here we analyze the situation further to relate the formation of
naked singularities or black holes with the initial data from which the 
collapse develops. The transition between the black hole and naked singularity 
phase is shown to be related to the nature of the initial data, and we also
examine how the local versus the global visibility of the naked singularity
changes with the change of essential parameters in the initial data.

In section 2 we briefly review the Szekeres model. In sections 3 and 4, 
we derive the necessary and sufficient condition for the existence of 
outgoing radial null geodesics (RNGs) from the singularity, and we relate 
the formation of naked singularities with the regular data specified on the 
initial hypersurface. In section 5, we investigate numerically the conditions 
for the global nakedness of the singularity in the TBL and Szekeres 
spacetimes. The final section 6 outlines some conclusions.

\section{Quasi-spherical gravitational collapse}

The metric in a comoving coordinate system has the form,
\be
ds^2=dt^2-M^2dr^2-N^2(dx^2+dy^2),
\ee
The energy-momentum tensor is of the form of an irrotational dust, 
\be
G^{ij} = T^{ij}= \epsilon \delta^{i}_{t} \delta^{j}_{t},
\ee
where $\epsilon (t,r)$ is the energy density of the cloud (here units are
chosen such that $c$ = $8\pi G$ = 1).  For convenience we adopt a pair
of complex conjugate coordinates, $\zeta =x+iy , \; \; \; \bar{\zeta}=x-iy,$
in which the metric takes the form
\be
ds^2=dt^2-M^2dr^2 - N^2d\zeta d\bar{\zeta}
\ee
Where M and N are general functions of $t,r,\zeta$ and
$\bar{\zeta} $.  The integrated form of Einstein's equations
is given by Szekeres\cite{Szeke},
\be
N={{R(r,t)}\over{Q(r,\zeta,\bar{\zeta)}}},~~~~ M={{QN'}\over{\sqrt{1+f}}},
\ee
restricted  to the case, $N'\equiv {{\partial N}\over{\partial r}}\ne 0$,
where the prime and dot denote partial derivatives with respect to
$r$ and $t$ respectively. Here $f(r)$ is an arbitrary function of $r$ subject 
to the restriction $f  > - 1,$ and $Q(r,\zeta ,\bar{\zeta})$ is of the form
\be
Q=a(r)\zeta\bar{\zeta}+B(r)\zeta +\bar{B}(r)\bar {\zeta}+c(r).
\ee
Here $a$ and $c$ are arbitrary real functions and $B$ is a complex
function of $r$, satisfying
\be
ac-B\bar{B}={\delta\over4}, \; \;  \delta=\pm 1  \; ,  \;  0.
\ee

We will consider here the case $\delta= +1$. In this case it is possible 
to reduce the two-metric by a bilinear transformation of coordinates
$\zeta,  \bar{\zeta}$ to the form,
\be
R^2(r,t)d\zeta d\bar{\zeta}/{1\over4}(\zeta\bar{\zeta}+1)^2,
\ee
which, on subsequent introduction of polar coordinates
$\zeta=e^{i\varphi}\cot{1\over2}\theta,~~ \bar{\zeta}=e^{-i\varphi}
\cot{1\over2}\theta,$
reduces (7) to the normal form of metric on a two-sphere given by
$R^2(r,t) (d\theta^2 + \sin^2\theta d\varphi^2).$ 
Hence, the two-surfaces $S_{r,t}$ defined by $t=const., r=const.$ 
are spheres of radius $R(r,t)$. Here $R(r,t)$ satisfies the  ``Friedmann 
equation'' 
\be
\dot R^2= f(r)+{{F(r)}\over R},
\ee
which is similar to the equation holding for the TBL models. Since we 
are studying collapse scenarios, we consider the collapsing branch of 
the solution, i.e. $ \dot{R} \le 0$. One more 
free function, corresponding to the epoch of singularity formation, 
arises when one integrates the above equation and we get
$$
\int (f(r) + \frac{F(r)}{R})^{-1/2}dR = t - t_0(r)
$$
Functions $F$ and $f$ are arbitrary functions of $r$, and $t_0(r)$ is 
another free function, which can be fixed using the scaling freedom. 
The density is given by,
\be
\rho={{QF'-3FQ'}\over{R^2(QR'-RQ')}}.
\ee

Therefore, the singularity curve is given by
\be
R(r,t_0) = 0, \; \; \; \; (t_0 > t_i),
\ee
\be
Q(r,t_{sc},\zeta)R'(r,t_{sc})-R(r,t_{sc})Q'(r,t_{sc},\zeta) = 0, \; \; (t_{sc} 
> t_i).
\ee
Here $t_i$, $t_0$ and  $t_{sc}$ give the times corresponding to that of
the regular initial hypersurface, and the shell-focusing and shell-crossing 
singularities respectively.

To analyze the shell-focusing singularity, it is 
essential to choose the initial data in such a way that $ t_i < t_0 < t_{sc}$.
The range of coordinates is given by
\be
R \neq 0, \; \; \; \; Q(r,t,\zeta)R'(r,t)-R(r,t)Q'(r,t,\zeta) \neq 0
\ee
We can easily check that if the functions $a(r)$, $B(r)$, $\bar B(r)$,
and $c(r)$ 
are constants then the equation (9) reduces to the corresponding
equation for the TBL metric. Amongst the functions $a$, $c$, $B$,
$\bar{B}$, $F$, $f$ and $t_0(r)$, equation (6) and a gauge freedom
would leave total five free functions in general. 
In comparison to the TBL models,
where we have only two free functions, namely $F(r)$ (the mass function) 
and $f(r)$ (the energy function), this model is clearly functionally more 
generic, allowing for a certain mode of non-sphericity to be included in the
consideration.

\section{Gravitational collapse and singularity formation}

Assuming the validity of Einstein's equations, the singularity 
theorems\cite{HawElli} ensure the formation of singularity in
gravitational collapse if some reasonable conditions are satisfied by
the initial data. For analyzing the detailed structure of the singularity 
curve, integrating equation (8) we get
\be
t - t_{0}(r) = - {{R^{3/2} G(-fR/F)}\over{\sqrt {F}}},
\ee
where $G(y)$ is a strictly convex, positive function  having the range $ 1 
\ge y \ge - \infty $ and is given by
\be   
G(y) =    \cases { {{\rm arcsin}\sqrt{y}\over y^{3/2}} - 
                     {\sqrt{1-y}\over y}, &  $1\geq y> 0$,\cr
                     {2\over 3}, & $y=0$,\cr
                     {-{\rm arcsinh}\sqrt{-y}\over (-y)^{3/2}} - 
                     {\sqrt{1-y}\over y}, & $0> y\geq -\infty$.\cr} 
\ee
Here $t_{0}(r) $ is a constant of integration. Using the scaling freedom, 
we can choose 
\be
R(0,r)=r,
\ee
and this gives
\be
{t_{0}(r)=}{{r^{3/2} G(-fr/F)}\over {\sqrt {F}}},
\ee
i.e. the singular epoch is uniquely specified in terms of the other free 
functions. The function $t_{0}(r)$ gives the time at which the physical radius 
of the shell labeled by $r =$ const.  becomes zero, and hence the shell 
becomes singular.

The quantity $R'$ which is useful for the further analysis, can be expressed 
using equations (13-16) as
\be
\begin{array}{clcr}
R'= &r^{\alpha - 1}[(\eta - \beta) X + [\Theta - (\eta - \frac{3}{2}\beta)
X^{3/2} G(-PX)]\times[P + \frac{1}{X}]^{1/2}]\\
~~~~\equiv &r^{\alpha - 1}H(X,r),
\end{array}
\ee
where we have used
$$
u = r^{\alpha}, \; X=(R/u),  \; \eta(r) = r\frac{F'}{F}, \; \; \beta(r) =
r\frac{f'}{f}, \; \; p(r) = r\frac{f}{F}, \; \; P(r)=p r^{\alpha-1}, \; \; 
$$
\be
~~~\Lambda = \frac{F}{r^{\alpha}}, \; \; \; \Theta \equiv \frac{t'_s \sqrt
\Lambda}{r^{\alpha-1}} =
\frac{1+\beta-\eta}{(1+p)^{1/2}r^{3(\alpha-1)/ 2}} +
\frac{(\eta - \frac{3}{2}\beta)G(-p)}{r^{3(\alpha-1)/ 2}}.
\ee
The function $\beta (r)$ is defined to be zero when $f = 0$.
All the quantities defined
here are such that while approaching the central singularity at $r = 0$
they tend to finite limits, thus facilitating a clearer 
analysis of the singular point. The factor 
$r^{\alpha}$ has been introduced for analyzing geodesics near the singularity.
The constant $\alpha (\ge 1)$ is determined uniquely using the condition 
that $\Theta(X,r)$  goes to a nonzero finite value in the limit $r$
approaching $0$ along any $X$ = constant direction. 
The angular dependence in the metric appears through $\xi,\bar\xi$ and
along radial null geodesics $\xi\bar\xi=$const. 

Using the metric (3), the equation for the RNGs is,
\be
{{dt}\over{dr}} =  {{QR'-RQ'}\over{Q\sqrt{1+f}}}.
\ee
Which in terms of $R$ and $u = r^\alpha$ can be rewritten as 
$$
{{dR}\over {du}}={{1}\over {\alpha r^{\alpha -1}}} {\biggr
[}\dot R {{dt}\over {dr}} +R' {\biggr ]}
={{\biggr [} {1-{{\sqrt {f+\Lambda /X}}\over {{\sqrt
{1+f}}}}(1-{{RQ'}\over{QR'}}){\biggr ]}} {{H(X,u)}\over {\alpha}}},~~~~~~~~~~~~~~~
$$
\be
~~~~~~~~\equiv U(X,u)+{{H(X,u)}\over{\alpha}}{{\sqrt {f+\Lambda /X}}\over 
{{\sqrt{1+f}}}}{{RQ'}\over{QR'}}.
\ee
For spherically symmetric models we get 
$Q' = 0$, and the above equation reduces to 
the TBL case. The  point $u = 0$, $R = 0$ is a singularity of the above 
first order differential equation. 
To study the characteristic curves of the above differential 
equation, we define $X=R/r^{\alpha}=R/u $. For the 
outgoing RNGs, $dR/du>0$ is a 
necessary condition. If future directed null geodesics meet the first 
singularity point in the past with a definite value of the tangent ($X=X_0$),
then using equation (20) and l'Hospital's rule, we can write for that value 
$X_{0} $,
\be
X_{0}=\lim_{R\to0,u\to0} {{R}\over {u}} =\lim_{R\to0,u\to0}
{{dR}\over {du}}=\lim_{R\to0,u\to0} {{\biggr [} {1-{{\sqrt 
 {f+\Lambda /X}}\over {{\sqrt
{1+f}}}}(1-{{RQ'}\over{QR'}}){\biggr ]}} {{H(X,u)}\over {\alpha}}}.
\ee
Since $Q(0) \ne 0$, $Q'(0)=0$  and near the central
singularity $R'\propto R/r$, if there is a definite tangent to the RNGs 
coming out, we can neglect $(RQ')/(QR')$ near the first singular point. 
Therefore the result is similar to the TBL model\cite{JosDwi},
\be
X_0=\lim_{{r\to0},{u\to0}}U(X,r)=U(X_0,0).
\ee
We have introduced here the notation that a subscript zero on any
function of $r$ denotes its value at $r = 0$. Defining 
\be
V(X)=U(X,0)-X={{\biggr [ }  {1-{{\sqrt {f_{0}+\Lambda_{0} /X}}\over {{\sqrt
{1+f_{0}}}}}{\biggr ]}}} {{H(X,0)}\over {\alpha}}-X,
\ee	
the existence of a real positive value $X = X_{0}$ for the above 
equation such that 
\be
V(X_{0})=0,
\ee
gives a necessary condition for the singularity to be at least locally naked.

In the neighborhood of the singularity we can write $R=X_{0}r^{\alpha}$, 
where $X_{0}$ is positive real root of above equation. To check whether the 
value $X_0$ is actually realized along any outgoing singular geodesic, 
we consider the equation for RNGs in the form 
$u = r^{\alpha} = u(X) $. From equation (20) we have 
\be
{{dX}\over{du}}={1\over{u}}{{\bigg [}{{{dR}\over{du}}-X}{\bigg ]}}
={1\over {u}}{{\bigg [}{U(X,u)-X + 
({{\sqrt{f+\Lambda/X}}\over {\sqrt{1+f}}}) 
{{X Q' u^{1/\alpha}}\over {\alpha Q}}} {\bigg ]}}.
\ee
The third term on the right hand side goes to zero faster  than
the first near $r = 0$. The solution of the above equation gives the
RNGs in the form $u = u(X)$. Now if the equation $V(X) = 0$ has $X=X_0$ 
as a simple root, then we can write
\be
V(X)\equiv(X-X_0)(h_0-1)+h(X),
\ee
where
\be
h_0={1\over{H_0}}{{\bigg [}{{{\Lambda_0H_0^2}\over{2\alpha
X_0^2\sqrt{f_0+\Lambda_0/X_0}\sqrt{1+f_0}}}+{{X_0N_0}\over{\sqrt{
f_{0}+\Lambda_0/X_0}}}}{\bigg ]}},
\ee
is a constant and the function $h(X)$ is chosen in such a way that 
\be
h(X_0)={{\bigg [}{{dh}\over{dX}}{\bigg ]}}_{X=X_0}=0,
\ee
i.e. $h(X)$ contains higher order terms in $(X-X_0)$ and
$H_0=H(X_0,0)$, $N_0=N(X_0,0)$ with $N(X,r)=-\dot R'r$.

As we are studying RNGs, $Q$ and $Q'$ are just the functions of $r$ 
(i.e. $u$),  therefore 
the equation for $dX/du$ can be written as
\be
{{dX}\over{du}}-(X-X_0){{h_0-1}\over u}={S\over u}
\ee
Here the function
\be
S=S(X,u)=U(X,u)-U(X_0)+h(X)+ {{\sqrt{f+\Lambda/X}}\over{\sqrt{1+f}}}
({{X Q' u^{1/\alpha}}\over {\alpha Q}})
\ee
is such that $S(X_0,0)=0$. Note that unlike the TBL case, here $S$ also 
has some functional
dependence on $\zeta$ and $\bar\zeta$. We can take care of
this by considering radial geodesics, i.e. only those with
fixed $\zeta$ and $\bar\zeta$.
Multiplying equation (29) by $u^{-h_0+1}$ and integrating we get
\be
X-X_0=Du^{h_0-1}+u^{h_0-1}\int Su^{-h_0}du,
\ee
where $D$ is a constant of integration that labels different geodesics.
We see from this equation that the RNG given by $D = 0$ always terminates
at the singularity $R = 0,u = 0$, with $X = X_0$ as the tangent. Also,
for $h_0 > 1$, a family of outgoing RNGs terminates
at the singularity with $X = X_0$ as the tangent. Therefore the 
existence of a real 
positive root to equation (24) is also 
a sufficient condition for the singularity 
to be at least locally naked.

Near the singularity the geodesic equation can be written as
$$
X-X_0 = Du^{h_0 -1}.
$$
Hence for $ h_0 >1 $ there is an infinite family 
of RNGs which terminate at the singularity in the past with $X_0$ as a
tangent. For $h_{0} \le 1 $ there is only one RNG corresponding to
$D=0$ which terminates at the singularity with $X_0$ as a tangent.
In ($R$,$u$) plane
$$
R-X_0 u= Du^{h_0},
$$
and for $h_0 > 1$ we have a family of infinitely many 
outgoing RNGs meeting the 
spacetime singularity in the past. For $h_0 \leq 1$ only $D = 0$ geodesic 
terminates at the physical singularity ($t_0(0), u = 0 $) in the past.

A relevant question is, if only one RNG has the tangent $X=X_0$ in the 
(u,R) plane then what is the behaviour of
the RNGs which are inside this root and very close to the singularity? They 
cannot 
cross the root so they have to start from the singularity in the past. And if 
they are outside the apparent horizon, then certainly the 
only possible tangent 
they can have at the singularity is the same as the tangent made by the 
apparent 
horizon at the singularity. To check if our differential equation for RNGs can 
satisfy this near the singularity, we note that in equation (20) the term in
the bracket goes to zero, while $H$ blows up in this limit. 
In the limit to the 
singularity, we expect $R/r^3$ to be $F_0$ along the RNG. To the lowest 
order we assume that along the RNG 
$$
R=F+M_{\delta}r^{\delta + 3}
$$ 
and check for the consistency by determining $M_{\delta}$ and $\delta$ in the 
limit. For the $f = 0$ case, using l'Hospital's rule we get,

$$
F_0 =\lim_{R\to0,r \to 0} {{R}\over {r^3}} =\lim_{R \to 0,r \to 0}
{{dR}\over {d(r^3)}}=
\lim_{R\to0,r\to0}{{\bigg [}1-\sqrt{F/R}{\bigg ]}}
{\bigg [} X+{1\over {\sqrt X}} {{nF_n}\over{F_0}} r^{n-3}{\bigg ]}
$$
\be
\; \;~~~~~~ =\lim_{R\to0,r\to0} {{1\over 6} {{nM_{\delta}F_n 
\over{{F_0}^{5/2}}}r^
{\delta +n-3}}},
\ee   
where $X =R/r^3$.
This leads to,
$$
\delta=3-n \; ~and~ \; M_{\delta}=
 -{{6 {F_0}^{7/2}}\over {nF_n}}.
$$
That means that our assumption is consistent. One can also check 
that any other 
kind of behavior of the RNGs near the singularity is not possible.

\section{End state of quasi-spherical dust collapse}

In this section, we analyze the end state of quasi-spherical dust
collapse evolving from a given regular 
density and velocity profile. Our purpose 
here is to determine how a given initial profile influences and characterizes
the outcome of the collapse. We will be
mainly studying the marginally bound ($f = 0$) case for convenience. The 
results can be generalized to non-marginally bound case using an analysis 
similar to that developed by Jhingan and Joshi\cite{JhiJoshi} for the
case of TBL models.

The initial density profile for a fixed radial direction ($\zeta$ and 
$\bar{\zeta}$ are constant) is given by $\rho=\rho(r)$, and if we are  given 
the function $Q$ then it is possible in principle to find out the
function $F(r)$ from the density equation (9). In order to give a physically 
clear meaning to our results, we assume all the functions (i.e. $\rho$, $Q$, 
and $F$ ) to be expandable in $r$ around the $r = 0$ point on the 
initial hypersurface. This means that even though the
functions are expandable, we do not necessarily require them to be $C^\infty$
and smooth. Therefore we take
\be
\rho=\sum_{n=0}^{\infty} \rho_n r^n,
\ee
\be
Q=\sum_{n=0}^{\infty} Q_n r^n ~~~~~~~~Q_0\ne0,
\ee
where the regularity condition implies $Q_{1} = 0$ \cite{Szeke}, and
\be
F=\sum_{n=0}^{\infty} F_n r^{n+3}.
\ee
Therefore
\be 
Q'=\sum_{n=2}^{\infty} n Q_n r^{n-1},
\ee
and,
\be
F'=\sum_{n=0}^{\infty}(n+3) F_n r^{n+2}.
\ee
Substituting the values of these functions in the density equation and 
matching the coefficients of different powers of $r$ on both the sides,
we get the coefficients of function $F(r)$ as,
$$ F_0={{\rho_0}\over 3}, ~~~~F_1={{\rho_1}\over4},
~~~~F_2={{\rho_2}\over5},
~~~~F_3={{\rho_3}\over6}-{{\rho_1}\over 12}{{Q_2}\over{Q_0}}$$
\be
F_4={{\rho_4}\over 7}-{{1}\over{7Q_0}}{\bigg {(}}{3\over 4}
\rho_1 Q_3 +{4\over 5}\rho_2 Q_2 {\bigg {)}}, 
~~~~F_5 ={{\rho_5}\over 8}-{1\over 8Q_0}{\bigg {(}}\rho_1 Q_4 
-{6\over5} \rho_2 Q_3 - \rho_3 Q_2 {\bigg{)}}.
\ee

We note that $Q$ has to obey the constraint given by equation (5) and 
(6) and $F(r)$
does not have any angular dependence. This means that the angular dependence
of different terms on right hand side of the equation (38) should exactly
cancel each other. This leads to the conclusion that $\rho_0$, $\rho_1$,
$\rho_2$ cannot have any angular dependence, and in general for $n \ge 3$,
$\rho_n$ to have some angular dependence  at least $\rho_{n-2}$ has to be
non-zero. This confirms that local nakedness behaviour in Szekeres model is 
essentially similar to the TBL model.

We now determine $\alpha $ and $\Theta_0$ which enter the 
equation for roots described in the previous section. From
$\eta = r F'/F$   we get $\eta$ as 
\be
\eta (r)=3+\eta _1r+\eta _2r^2+\eta _3r^3+O(r^4),
\ee
where $\eta_1,\eta_2,\eta_3, . . . .$ are constants related to $F(r)$.  
It turns out that we need the explicit form of the terms up to
order $r^3$ only. If all the derivatives $\rho_n$ of the density
vanish, for $n \le (q-1)$, and the $q^{th}$ derivative is the first
non-vanishing derivative, then $T_{\eta}^q $, the $q^{th}$ term in
the expansion for $\eta $ is,
\be
T_{\eta}^q={{qF_q}\over {F_0}}r^q.
\ee
Here $q$ takes the value 1, 2 or 3. In this case,
\be
\eta (r)=3+{{qF_q}\over{F_0}}r^q + O(r^{q+1}).
\ee
From equation (18), we get for marginally bound cloud ($f=0$),
$\beta = p = P = 0$ and using $\eta$ from equation (41) we get,
\be
\Theta={{3-\eta}\over{3r^{3(\alpha-1)/2}}}=-{{qF_q}\over{3F_0}}
{{r^q}\over{r^{3(\alpha-1)/2}}},
\ee
correct up to order q in $\eta$. The constant $\alpha$ is
determined by the requirement that $\Theta_0$ is finite and
non-zero as $r\to0$, which gives,
\be
\alpha=1+{{2q}\over{3}}, ~~~~ \Theta_0=-{{qF_q}\over{3F_0}}.
\ee
The root equation (23) with $f = 0$ and $\eta_0=3$, reduces to
\be
V(X)={1\over{\alpha}}{{\biggr [}{1-\sqrt{{{\Lambda_0}\over{X}}} 
{\biggr ]}}}{{\biggr [} {X+{{\Theta_0}\over{\sqrt X}}} {\biggr
]}}-X , ~~~~ V(X_0)=0,
\ee
with $\Theta_0$ given by equation (43). The limiting value of
$\Lambda =F/r^{\alpha}$ here is ,
\be
{\Lambda_0 =\cases{ 0,&$ q<3$,\cr
			F_0,&$ q=3$,\cr
			\infty ,&$ q>3$.\cr}}
\ee
As $\Lambda_0$ takes different values for different choices of q, the 
nature of the roots depends upon the first nonvanishing derivative of the 
density at the center.

This then leads to the conditions for local nakedness of the singularity 
in Szekeres models, similar to those obtained by Singh and Joshi
\cite{JosSin} for the TBL models. The main similarity comes due to the 
Friedmann equation which is same in both the Szekeres and TBL cases. 
We consider the cases below:

\subsection{ $\rho_1 \ne 0$ }

In this case, q=1, $\alpha=5/3$, and equation (44) gives,
\be
X_0^{3/2}=-{{F_1}\over {2F_0}}=-{3\over 8}{{\rho_1}\over{\rho_0}}.
\ee
We assume the density to be decreasing outwards, $\rho_1<0$,
hence $X_0$ will be positive and singularity is at least locally naked.

\subsection{$\rho_1=0 $ , $\rho_2\ne 0$ }

In this case, $q = 2$, $\alpha=7/3$, and equation (44) gives,
\be
X_0^{3/2}=-{{F_2}\over {2F_0}}=-{3\over {10}}{{\rho_2}\over{\rho_0}}.
\ee
Since density is decreasing outwards, i.e. $\rho_2<0$, this implies
$X_0$ will be positive and again singularity is at least locally naked.

\subsection{$ \rho_1=\rho_2=0, \rho_3\ne 0$} 
 
Now we have $q = 3$,  $\alpha =3 ,\Lambda_0=F_0$, 
$\Theta_0=-F_3/F_0= -\rho_3/2\rho_0 $, and $\rho_3<0 $. As in this case 
$\Lambda_0=F_0\ne 0$, the equation (44) for the roots becomes,
\be
{1\over 3}{{\bigg (}{1-\sqrt{{F_0}\over X}}{\bigg )}} {{\bigg (}
{X-{{\rho_3}\over{2\rho_0\sqrt X}}}{\bigg )}}-X =0.
\ee
Substituting $X=F_0x^2$ and $\xi=F_3/F_0^{5/2}={3\sqrt 3} \rho_3
/2\rho_0^{5/2}$  gives
\be
2x^4+x^3+\xi x- \xi=0.
\ee
Now using the standard methods we can check for the possible  
existence of real positive roots for the above quartic. We see that for
the values $\xi<\xi_2=-25.99038$ real positive roots exist and we have 
naked singularity. In the case otherwise, we have collapse resulting 
into a black hole.

\subsection{ $\rho_1=\rho_2=\rho_3=0 $} 

In this case of $q \ge 4$ we have $ \alpha\ge11/3$, and
$\Lambda_0 =\infty $. So positive values of $X_0 $ cannot satisfy
the equation (44) for the roots, and the collapse always ends in a black hole 
(in the case of homogeneous collapse, $q=\infty$ and the collapse
leads to a black hole, i.e. the Oppenheimer-Snyder result is a
special case of our analysis).

\subsection{ $\alpha =0,~~~~\Lambda_0\ne0$ }
 
In this case  $f=0, F = r\Lambda , \Lambda_0 \ne 0$, and 
the assumption in (34) is not valid. Now from equations (17,18) we get, 
\be
\Theta_0={2\over 3},~~~~H(X,0)={1\over 3}X+{2\over{3\sqrt X}},
\ee
and equation (23) reduces to,
\be
x^4+{1\over 2} \sqrt{\Lambda_0}x^3-x+\sqrt{\Lambda_0}=0,
\ee
where $X=x^2$.

We look for the real positive roots of this quartic in the same way
as earlier. Substituting  $\beta = \Lambda_0^{3/2}/12 $, we get 
the condition for the real root,
\be
{\beta >\beta_1=17.3269~~~~or~~~~\beta <\beta_2=6.4126\times {10^{-3}}}.
\ee
But we can easily see that for the first case the above equation cannot be 
satisfied for any positive values of $x$. Therefore we get the condition for 
existence of naked singularity to be $\beta <\beta_2=6.4126\times
{10^{-3}}$, which is the same as the condition  obtained earlier by Joshi and 
Singh\cite{JosSingh}.

\section{Global visibility}

Amongst various versions of cosmic censorship available in the literature,
the strong cosmic censorship hypothesis does not allow the singularities 
to be even locally naked. The spacetime must then be globally hyperbolic.
Hence the case of existence of a real positive root to the equation (24) is a 
counter-example to strong cosmic censorship version. However, one may take 
the view that the singularities which are only locally naked may not be of
much observational significance as they would not be visible to observers
far away, and as such the spacetime outside the collapsing object may be
asymptotically flat.  On the other hand, globally naked singularities could 
have observational significance, as they are visible to an outside observer
far away in the spacetime. Thus, one may formulate a weaker version of 
censorship, whereby one allows the singularities to be locally naked but
rules out the global nakedness. In the present section, we now examine this
issue of local versus the global visibility of the singularity for the 
aspherical models under considerations, and also for the TBL models, which
provides some good insights into what is possible as regards the global
visibility when we vary the initial data, as given by the initial density
distributions of the cloud.

In fact, all outgoing geodesics which reach $r=r_c$ with $R(t,r_c) > F(r_c)$ 
will reach future null infinity. We assume $f(r)$ and $F(r)$ to be 
at least $C^2$ in 
the interval $0 < r \leq r_c$. Let us consider the situation when $V(X) = 0$ 
admits a real simple root $X=X_0$, and a family of outgoing RNGs terminate 
in the past at the singularity with this value of tangent. Since the 
geodesics emerge from the singularity with tangent $X_0$, and the apparent 
horizon has the tangent $\Lambda_0$ at the singularity, the local nakedness 
condition implies $X_0 > \Lambda_0$.

Rewriting $D$ in equation (31) in  terms of $X_c$ and $u_c$, where $X_c$ 
and $u_c$ are values of X and u on the boundary of the dust cloud, we have
\be
X-X_0=(X-X_c){{\bigg [}{u\over{u_c}}{\bigg ]}^{h_0-1}}+u^{h_0-1}
\int_{u_c}^u Su^{-h_0}du.
\ee
The event horizon is represented by the null geodesic which satisfies the 
condition $X_c=\Lambda (r_c)$. The quantity $dR/du$ is positive for the
outgoing geodesics from the singularity at $r = 0$. Therefore, all the null
geodesics which reach the line $r=r_c$ with $X_c > \Lambda (r_c)$, where the
metric is matched with the exterior Schwarzschild metric, will escape to 
infinity and the others will become  ingoing.  The matching of the Szekeres 
models with the exterior Schwarzschild spacetime is given by Bonner 
\cite{Bonn}.

Our method here to investigate the global nakedness of the singularity
will be to integrate the equation (20) for the RNGs numerically. For 
simplicity and clarity of presentation 
we will consider the case $f=0$ only. We expect that the overall 
behavior will be similar in $f\ne 0$ case also (for any given $\alpha$),
as evidenced by some test cases with $f\ne0$ we examined; however, we will 
not go into those details here. The mass function $F$ is taken to be 
expandable, and also we take the 
density to be decreasing outwards, as will be the case typically for any
realistic star, which would have a maximum density at the center.
There may be some possibility to study the behavior of RNGs near the 
singularity using analytic techniques, however, in a general case,
we do not have analytical solutions for the concerned differential equation. 
Hence, we use numerical techniques to examine the global behavior of the RNGs,
and to get a clearer picture, we keep only the first nonvanishing 
term after $F_0$ in the expansion of $F(r)$. Thus, we have,
$$ F= F_0r^3 + F_n r^{3+n},$$
i.e. for the TBL model,
$$\rho=\rho_0 +\rho_nr^n $$
where $F_n \le 0$. Hereafter we fix $F_0 =1$ which is in 
some sense an overall scaling. First we consider 
below the TBL model, and then some cases of Szekeres quasi-spherical 
collapse will be discussed to gain insight into the possible 
differences caused by the introduction of asphericity.

\begin{figure}[p]
\parbox[b]{7.88cm}
{
\epsfxsize=7.85cm
\epsfbox{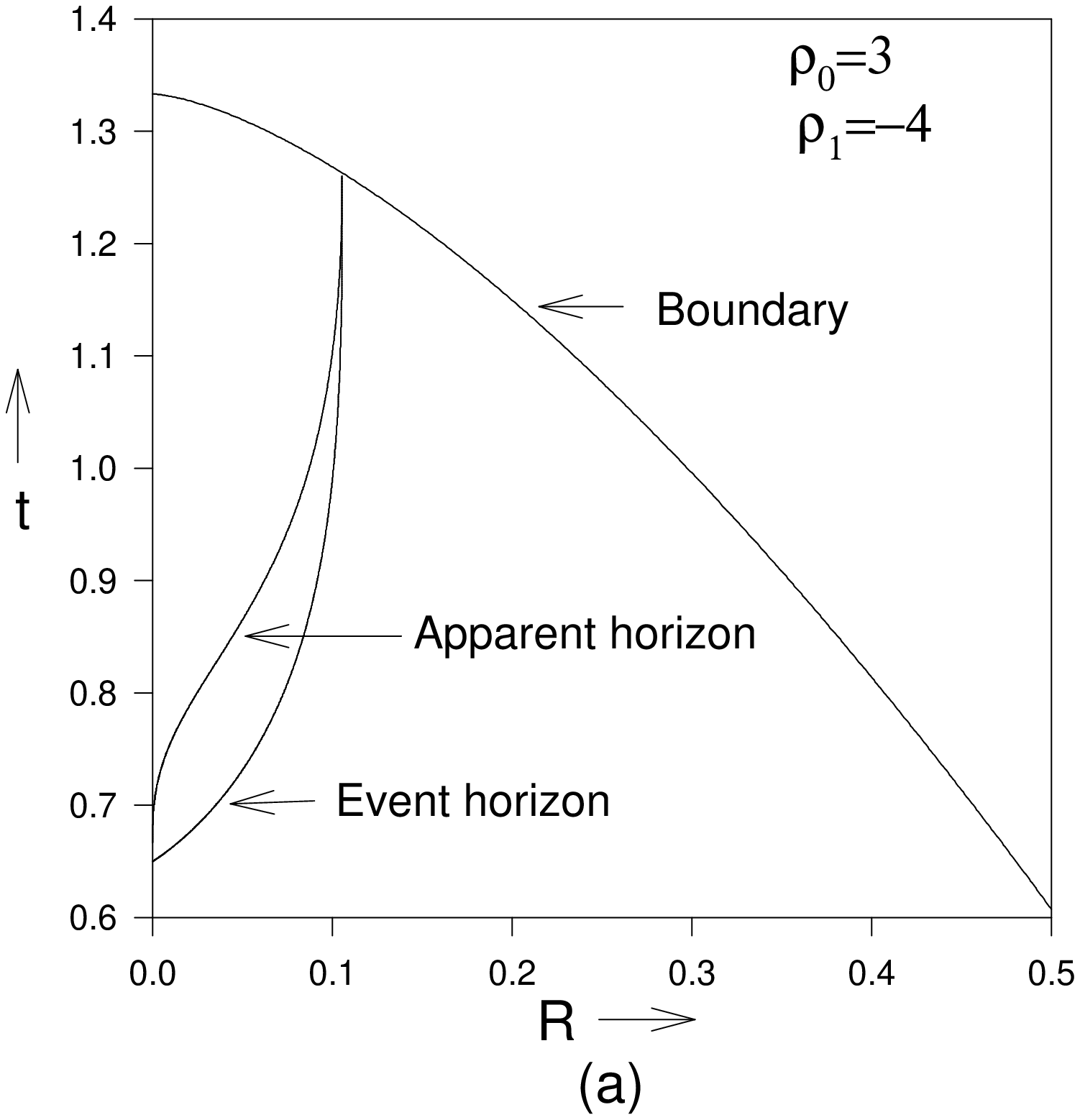}
}
\ \ \    
\parbox[b]{7.88cm}
{
\epsfxsize=7.85cm
\epsfbox{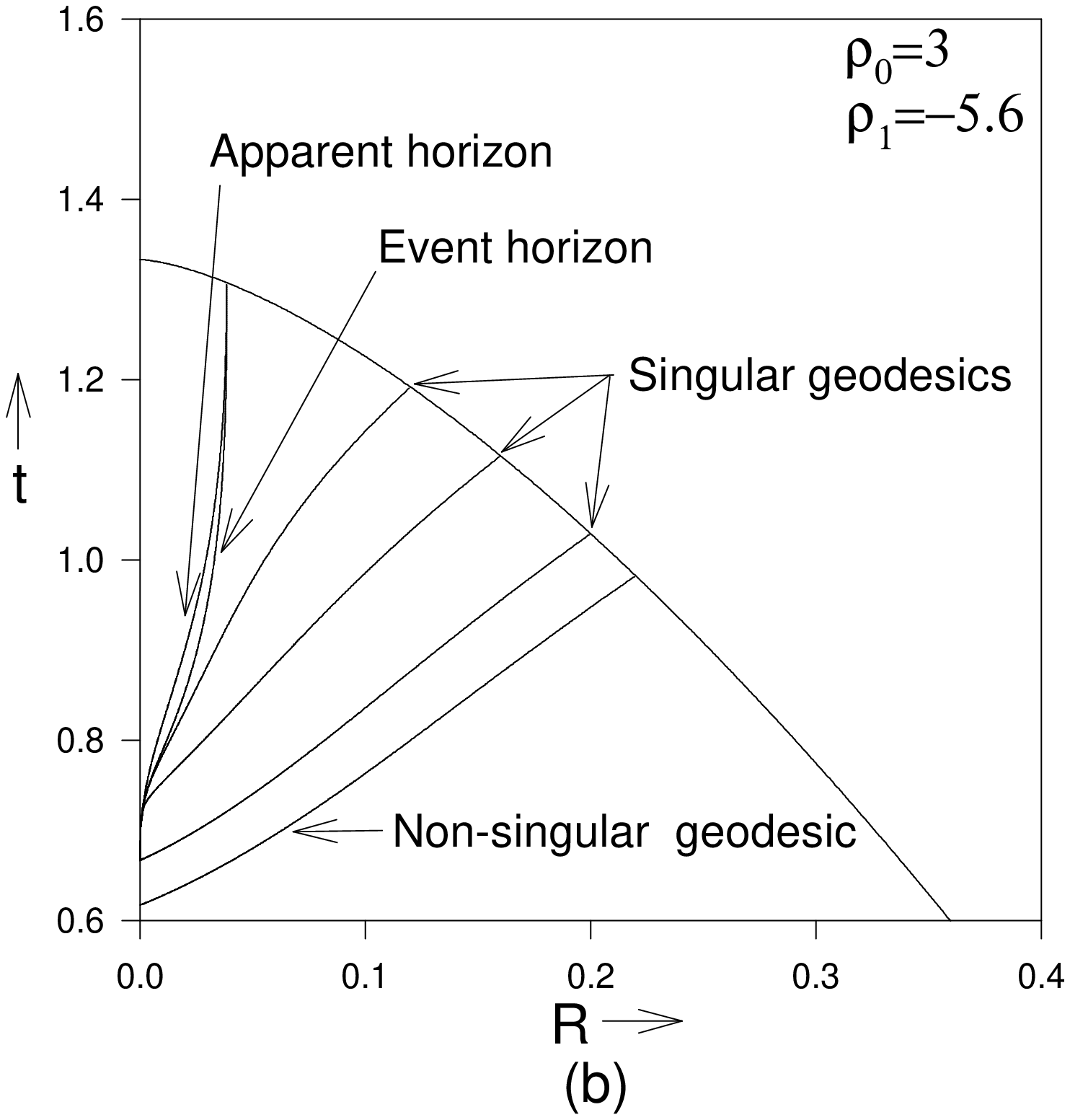}
}
\end{figure}
%\vskip 1.5cm
\begin{figure}[p]
\ \ \
\parbox[b]{7.88cm}
{
\epsfxsize=7.85cm
\epsfbox{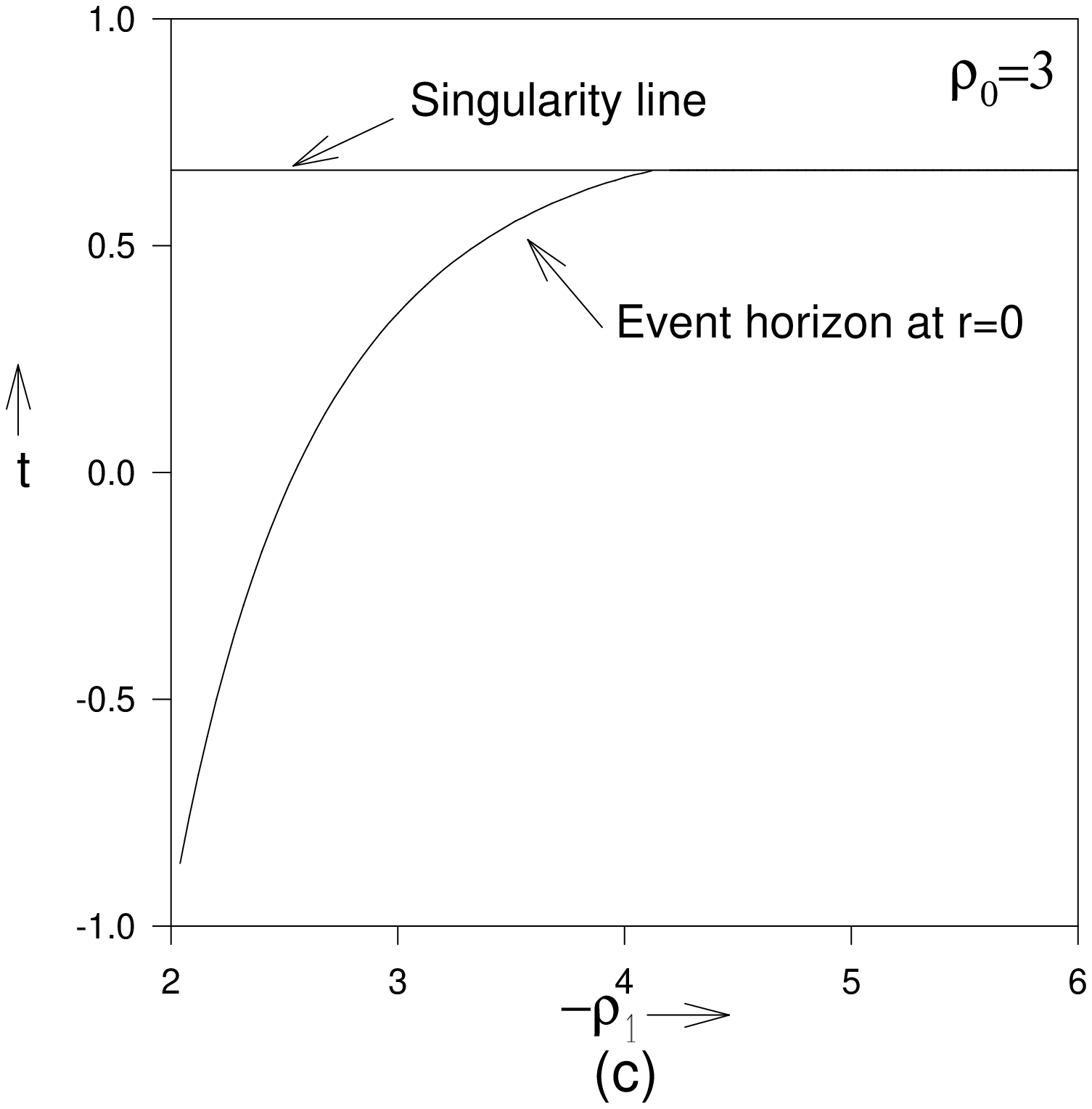}
}
\caption{
The structure of singularity in the marginally bound (f=0) 
TBL model for $\rho_1
\ne 0$ case. Here $\rho_0=3$, so the epoch of singularity formation is at 
$t=t_0(0)=2/3$. (a) The event horizon meets the center $r=0$ below the 
singularity and no null trajectories escape to infinity. The singularity is 
only locally naked.(b) Singularity is globally naked. There is a family of 
null geodesics from the singularity which meet the boundary of
cloud with $R_c>F_c$. (c) Graph showing transition from local nakedness to 
global nakedness, by varying $\rho_1$.}
\end{figure}

\subsection {TBL model }

Using the local nakedness conditions on the initial data as discussed
in the earlier section, we analyze here the global behaviour of RNGs in 
the TBL model.

\subsubsection {$F_1 \ne 0$}

In this case, the singularity is always locally naked, and we have a 
family of null geodesics coming out of the singularity with apparent horizon 
as the tangent at the singularity. We see in this case that the singularity
can be locally or globally naked (Fig. 1(a), 1(b)). This basically depends 
upon the chosen value of $\rho_1$, and where we place the boundary
of the cloud. When the singularity is globally naked, all the RNGs which meet 
the boundary of the cloud with $R< R_{crit}$ start from the singularity, and 
the other null geodesics start from the regular center $r=0$ before the 
singularity is formed.

If we match the density smoothly to zero at the boundary, then $r_c=-3F_0/4F_1
= -\rho_0/\rho_1$, and we see that for $F_1< -1.12$ i.e. $\rho_1 <-4.48$ the 
singularity is globally naked. In this case the event
horizon starts at the singularity $r=0,~~t=2/3$ as shown in Fig. 1(c). 
It is seen that for the other case the event horizon formation starts
before the epoch of formation of the singularity. It is thus seen that as
we depart away from the homogeneous density distribution (the Oppenheimer-
Snyder case), by introducing a small density perturbation by choosing a small
nonzero value of $\rho_1$, the singularity becomes naked. However, it will
be only locally naked to begin with 
and no rays will go outside the event horizon, thus
preserving the asymptotic predictability. It follows that even though the
strong version of cosmic censorship is violated, the weak censorship is 
preserved. It is only when the density perturbation is large 
enough (Fig. 1(c)),
beyond a certain critical value, then 
the singularity will become globally naked, thus becoming
visible to the outside observers faraway in the spacetime.

\vskip 1 truein
\begin{figure}[p]
\parbox[b]{7.88cm}
{
\epsfxsize=7.85cm
\epsfbox{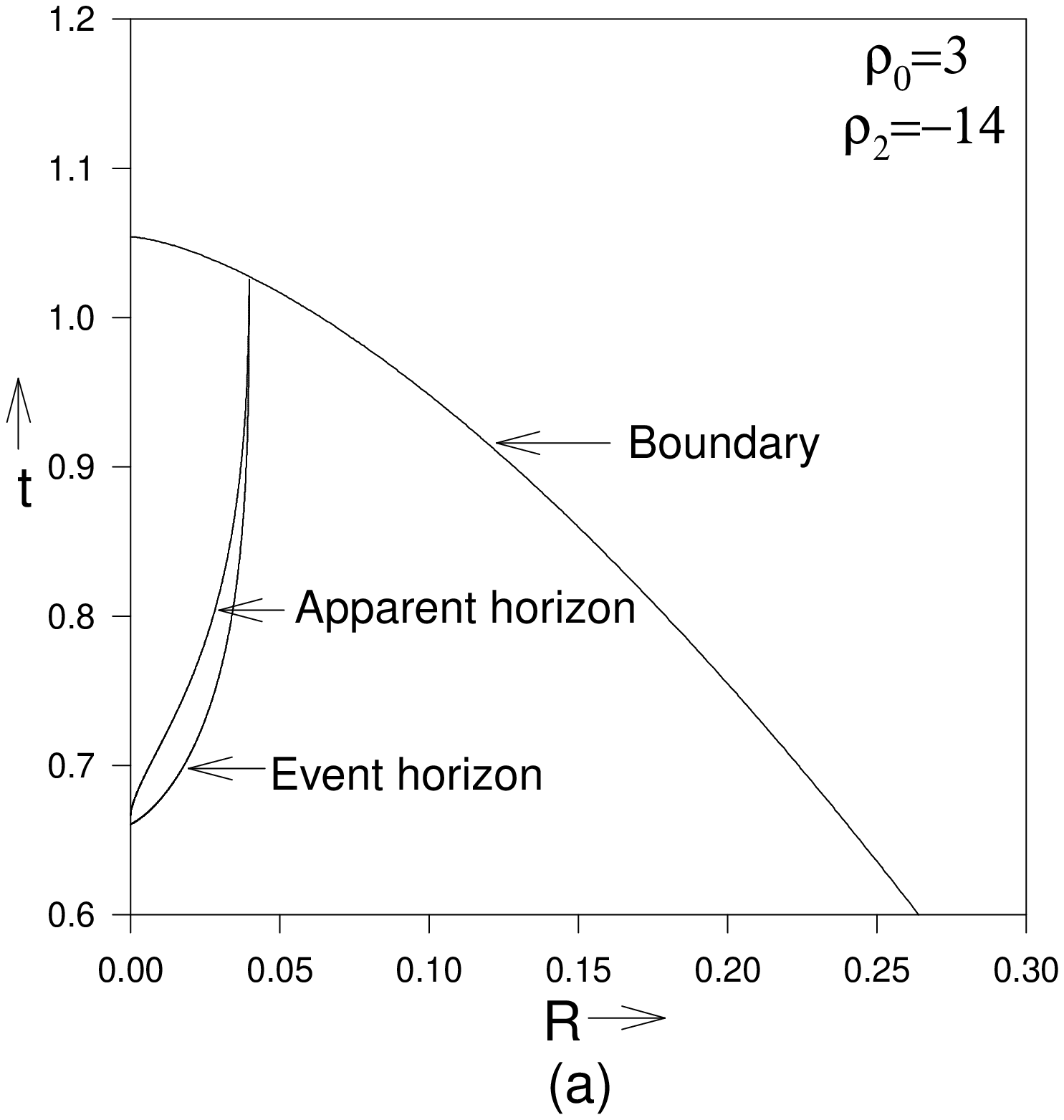}
}
\ \ \
\parbox[b]{7.88cm}
{
\epsfxsize=7.85cm
\epsfbox{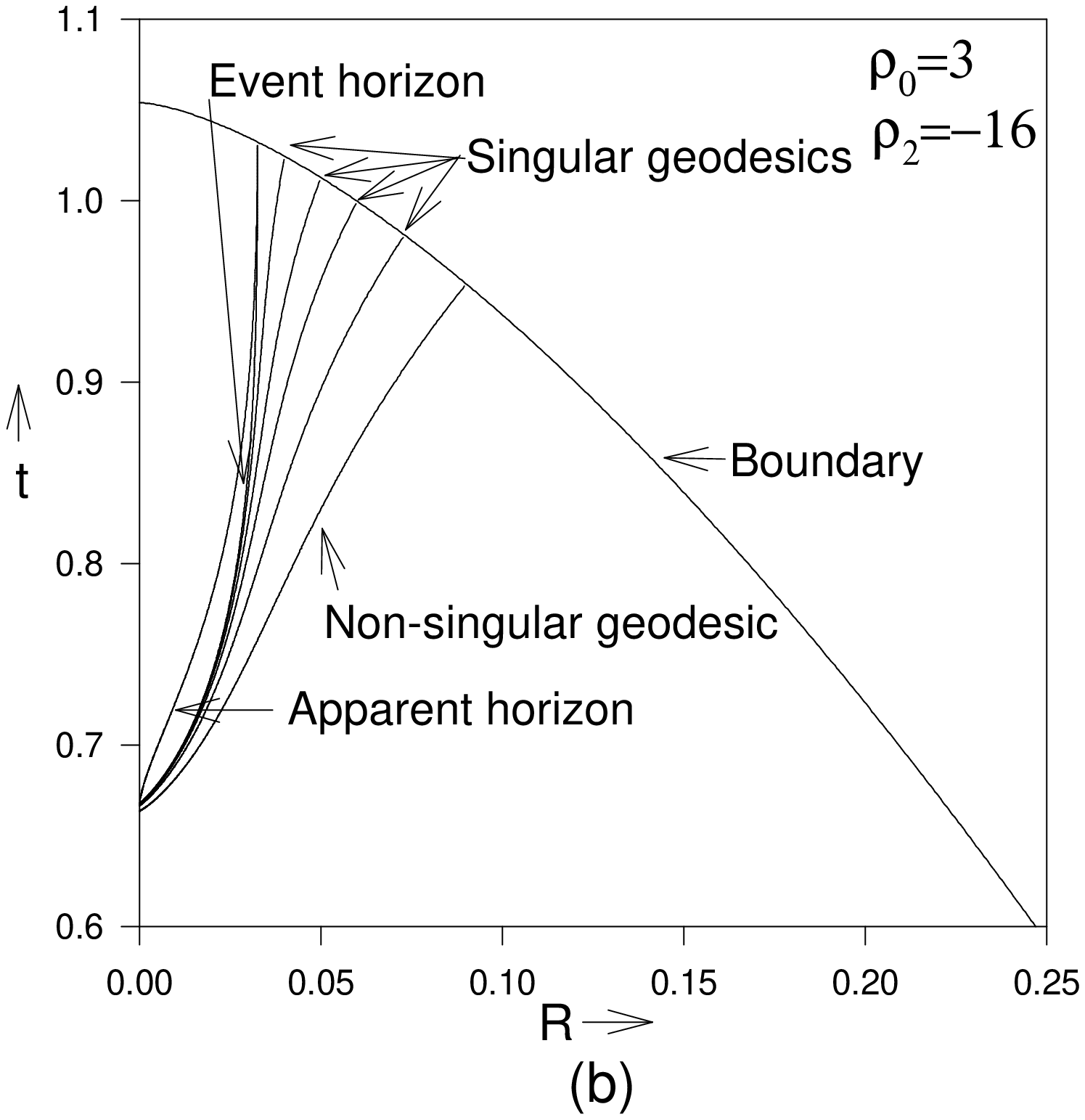}
}
\end{figure}
\begin{figure}[p]
\parbox[b]{7.88cm}
{
\epsfxsize=7.85cm
\epsfbox{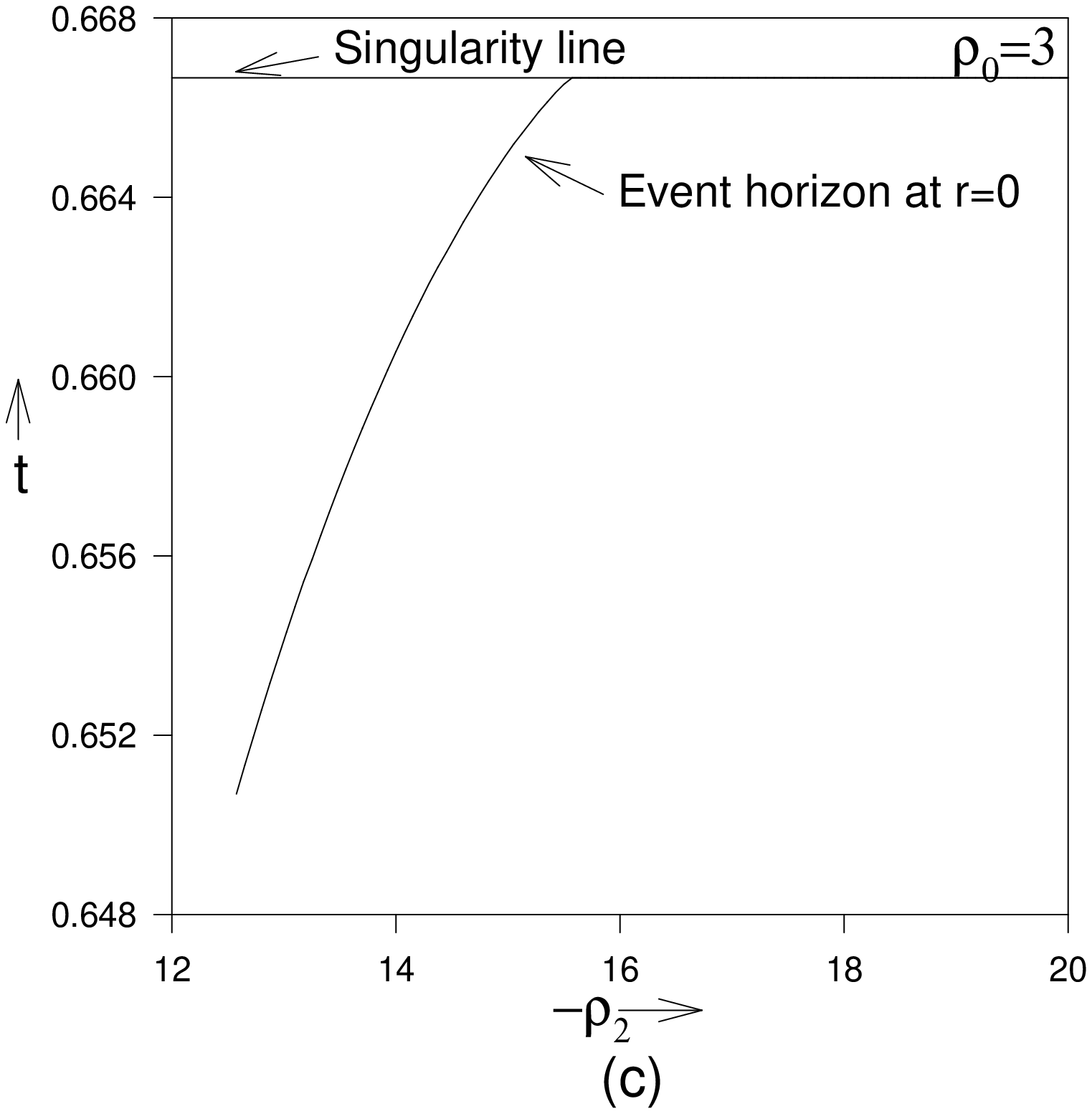}
}  
\caption{
The structure of singularity in marginally bound (f=0) TBL model for $F_2\ne 
0$, i.e $\rho_2\ne 0$  case. We have $\rho_0=3$, and $t_0(0)=2/3$. 
(a) Singularity is only locally naked. (b) Singularity is globally naked. 
There is a family of geodesics from the singularity which meets the boundary 
of the cloud with $R_c>F_c$. (c) The graph showing a transition from the 
locally naked to globally naked singularity by varying $\rho_2$.}
\end{figure}

\subsubsection {$F_2 \ne 0$}
	
The overall behavior in this case (Fig. 2(a), 2(b)) is very similar to 
$F_1\ne 0$ case discussed above. The singularity is always locally naked 
whenever $F_2$ is nonzero, and there is a family of RNGs coming out of 
singularity with apparent horizon as the tangent.
If we match the density smoothly to zero at the boundary of the cloud,
then $r_c=\sqrt {-3F_0/5F_2}=\sqrt{-\rho_0/\rho_2} $. We see that for $F_2< 
-3.1$, i.e. $\rho_2<-15.5$, the singularity is always globally 
naked, while in the other cases it is not globally naked (fig. 2c).

\vskip 1 truein
\begin{figure}[p]
\parbox[b]{7.88cm}
{
\epsfxsize=7.85cm
\epsfbox{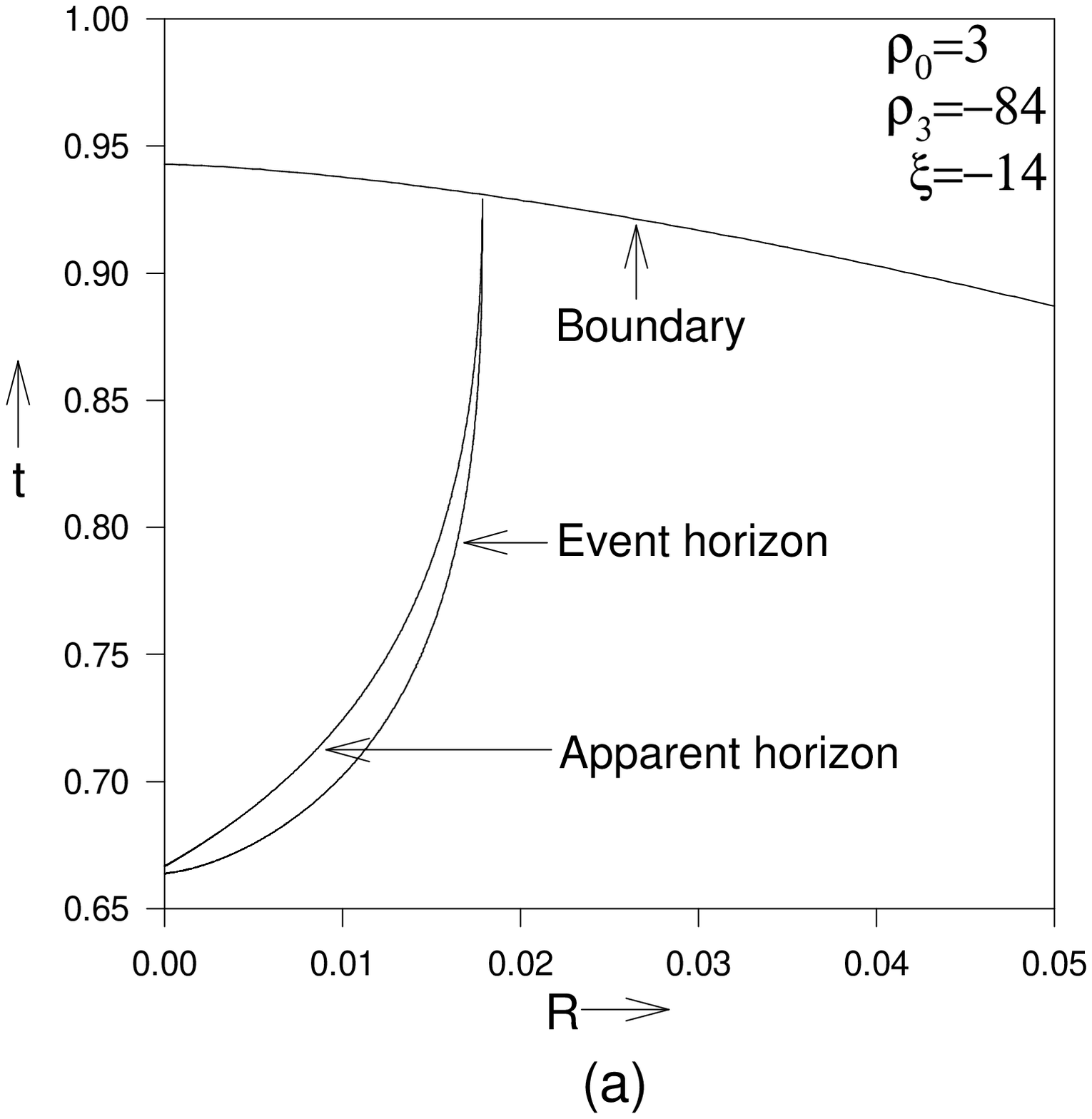}
}
\ \ \
\parbox[b]{7.88cm}
{
\epsfxsize=7.85cm
\epsfbox{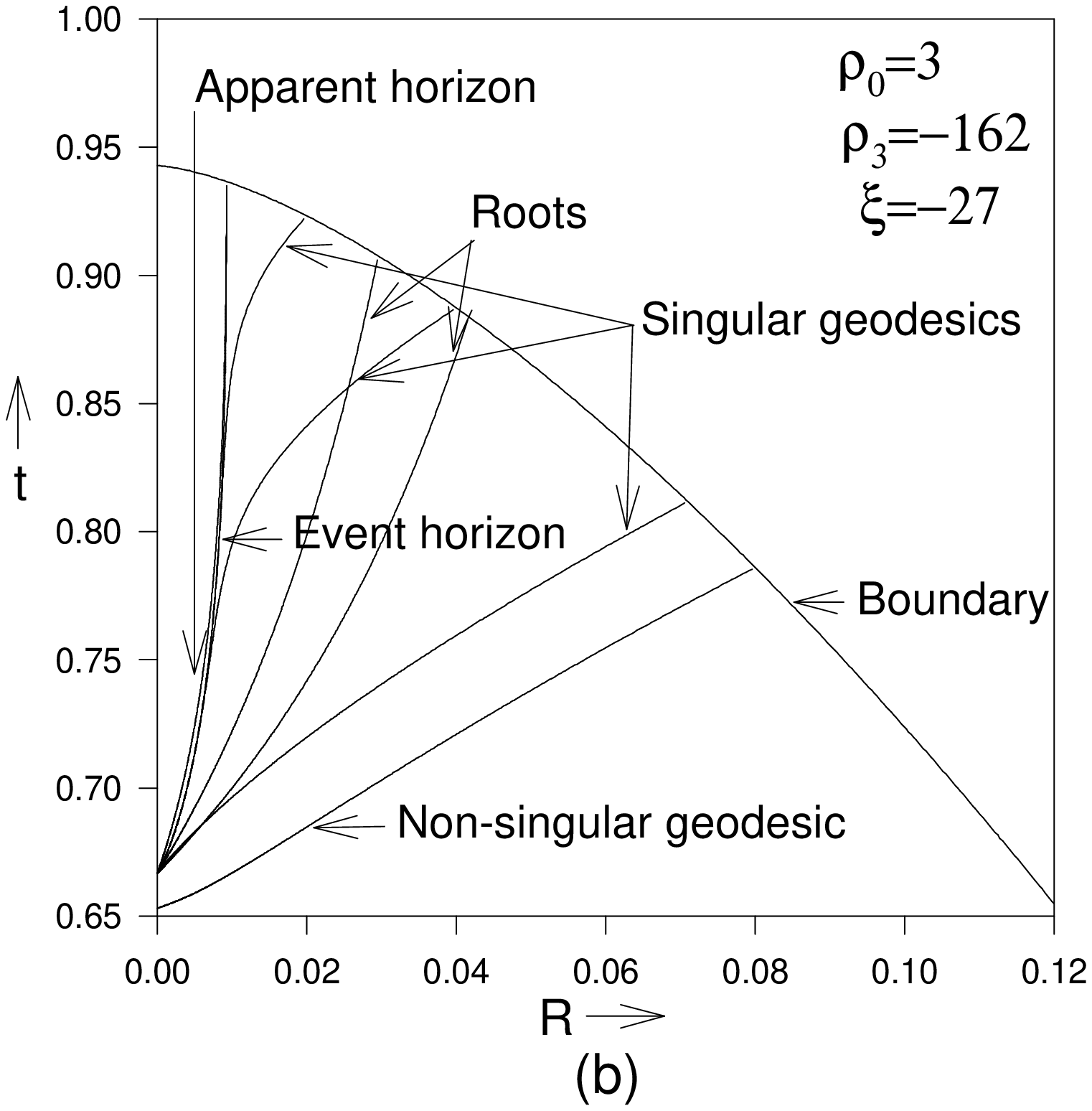}
}
\end{figure}
\begin{figure}[p]
\parbox[b]{7.88cm}
{
\epsfxsize=7.85cm
\epsfbox{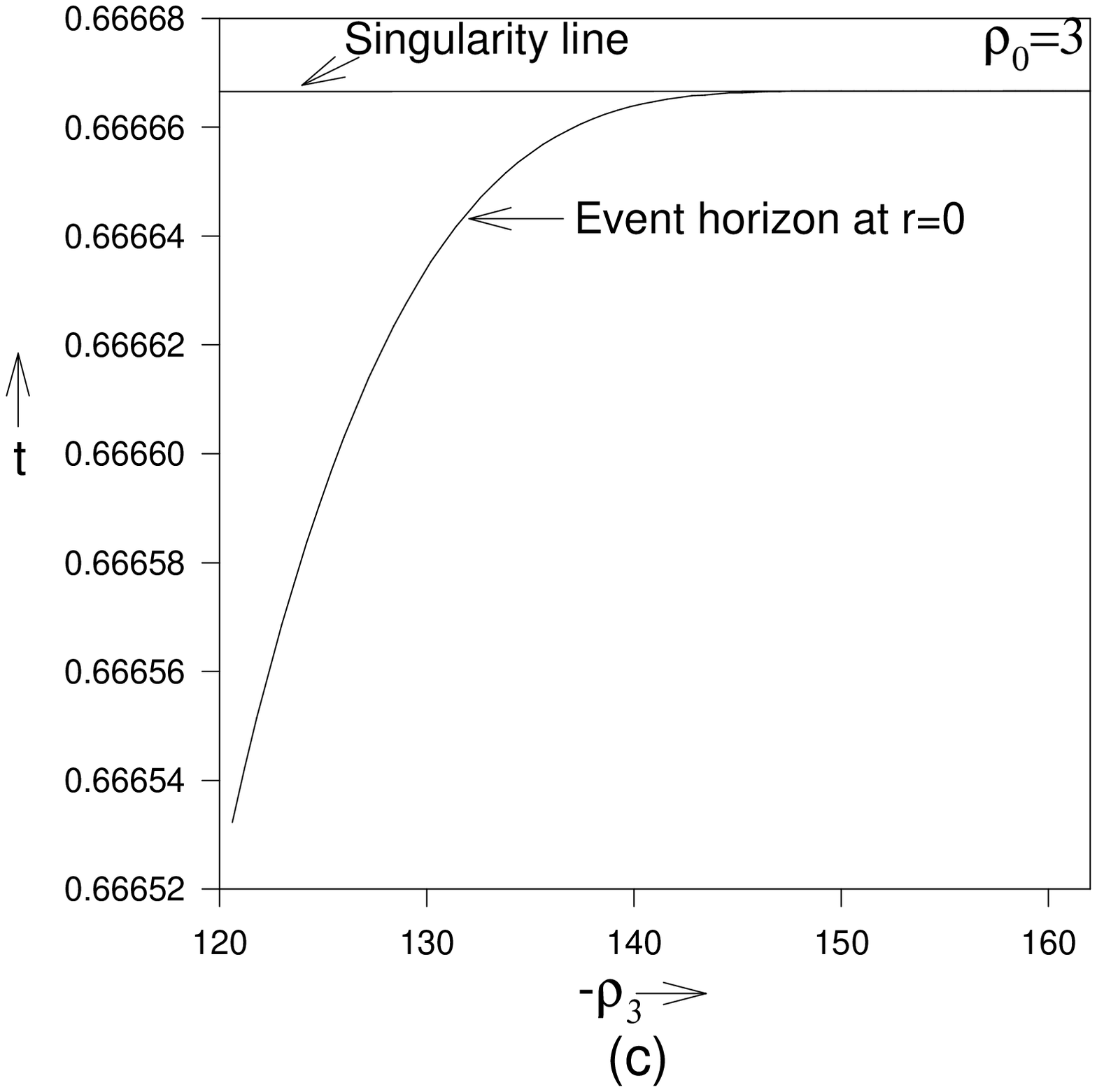}
}  
\caption{
Naked singularity in marginally bound (f=0) TBL model for $F_3\ne 0$, i.e. 
$\rho_3\ne 0$ case.  (a) Singularity is only locally
naked. (b) Singularity is globally naked. There is a family of geodesics from 
the singularity which meets the boundary of the cloud with $R_c>F_c$. 
(c) The graph showing transition from globally naked singularity to a 
black hole by varying $\rho_3$.}
\end{figure}
\subsubsection{$F_3 \ne 0$}

This is the case when the first two derivatives of the density, $\rho_1$
and $\rho_2$ vanish but $\rho_3\ne0$.
In this case $\alpha= 3$, and we have two real positive roots for 
the equation $V(X)=0$ if $\xi={F_3}/{F_0}^{5/2} < -25.99038=\xi_{crit} $. 
The singularity is then locally naked, and we see that it is 
also globally naked with a family of null geodesics meeting the singularity 
in the past with 
$X_0$ as the tangent (Fig. 3(b)), where $X_0$ is the smaller of the two roots.
In the range $-1.5 > \xi > \xi_{crit}$, the 
apparent horizon forms at the time of singularity but the singularity is not 
even locally naked. Numerically we see that (Fig. 3(c)) in this case the 
event horizon formation starts before the singularity formation as expected.

\subsection {Szekeres models}
	
The spherically symmetric TBL models involve no angular dependence for the 
physical variables such as densities etc.
However, it may be possible to examine the effects of such a dependence by
using the Szekeres models, which incorporate a certain mode of departure
from the spherical symmetry. Thus, this may be a good test case to examine
what is possible when non-sphericities are involved. With such a purpose
in mind, we keep the form of the mass function $F$ the same as that
for the TBL models 
discussed above, and as a test case we choose the form of $Q$ as given below,
$$
Q = \mathrm{cosh}(ar)\zeta \bar{\zeta} + 
\frac{1}{4 \mathrm{cosh}(ar)}.
$$
The constant $a$ is chosen in such a way that all regularity conditions are 
satisfied. Again, we consider the subcases corresponding to different
density distributions, as considered earlier.

\subsubsection {$F_1 \ne 0$}

The singularity is always locally naked and there is a family of RNGs with 
$R=F$ near the singularity. We choose a= 1.0. The general behavior of RNGs
is similar to that of the similar case in TBL model. But there is a new 
feature emerging now, which is that the global visibility of singularity 
in some critical regions can depend upon the value of $\zeta \bar{\zeta}$. 
That is, the singularity will be globally naked for a certain range of  
$\zeta \bar{\zeta}$, while for some other range of  $\zeta \bar{\zeta}$ 
it would not be globally naked with the event horizon starting to form 
before the singularity for these values of $\zeta \bar{\zeta}$ (Fig. 4a).

If we take the boundary of cloud at $r= r_c= -3F_0/4F_1$ then in the 
range $-1.06 <F_1 < -1.01$, the singularity is globally naked depending on the 
value of $\zeta \bar{\zeta}$ (Fig. 4a). Therefore the global visibility of 
singularity is directional. This feature distinguishes singularities in
Szekeres spacetimes from TBL model in that the global visibility of the
singularity will be related to the directional or angular dependence. 
For the values $F_1< -1.06$, the singularity is globally visible independent 
of the direction i.e. the values of $\zeta \bar{\zeta}$, whereas for 
$F_1> -1.01$ the singularity is not globally visible for any $\zeta 
\bar{\zeta}$ values. 

\vskip 1 truein
\begin{figure}[p]
\parbox[b]{7.88cm}
{
\epsfxsize=7.85cm
\epsfbox{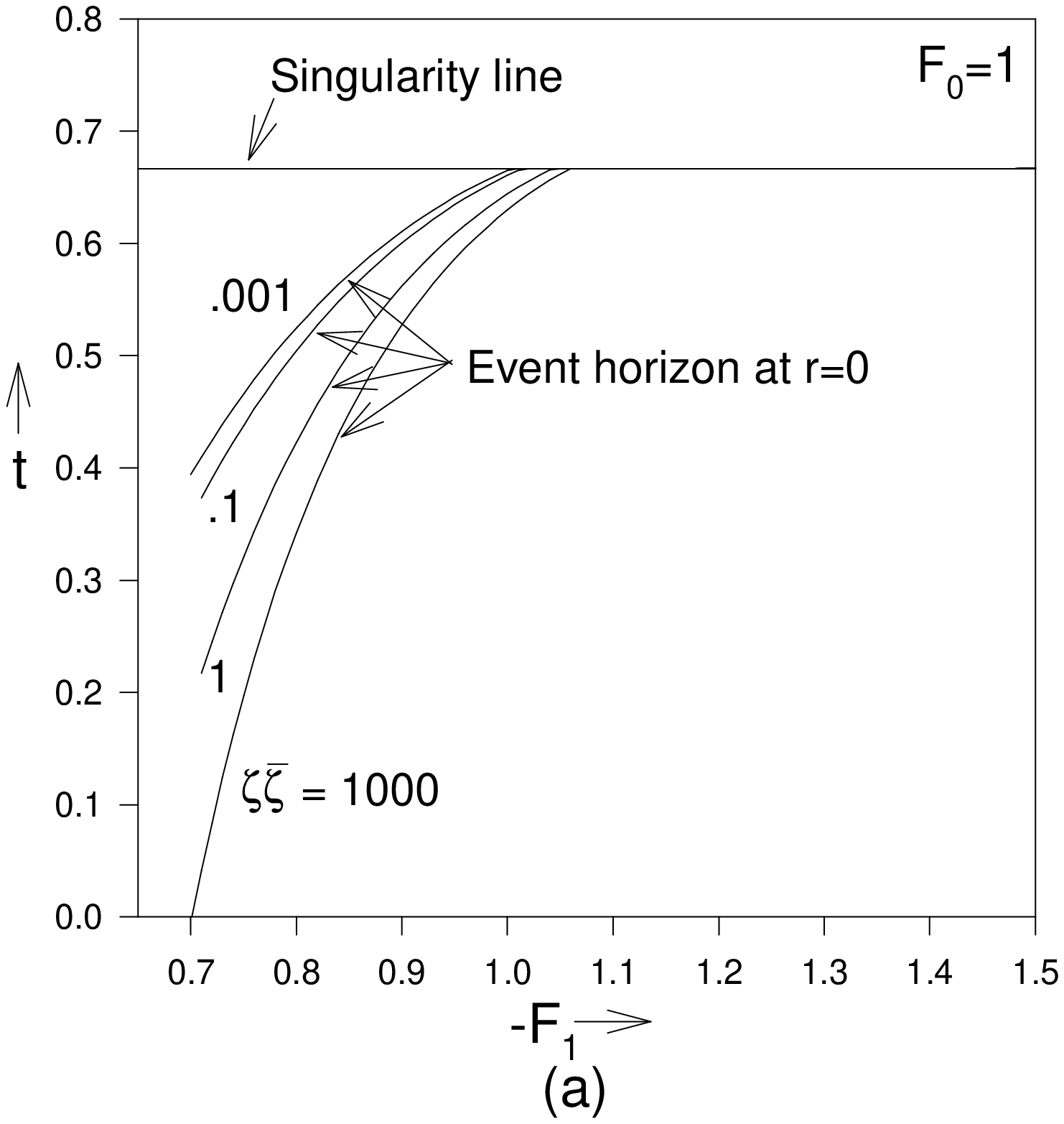}
}
\ \ \  
\parbox[b]{7.88cm}
{
\epsfxsize=7.85cm
\epsfbox{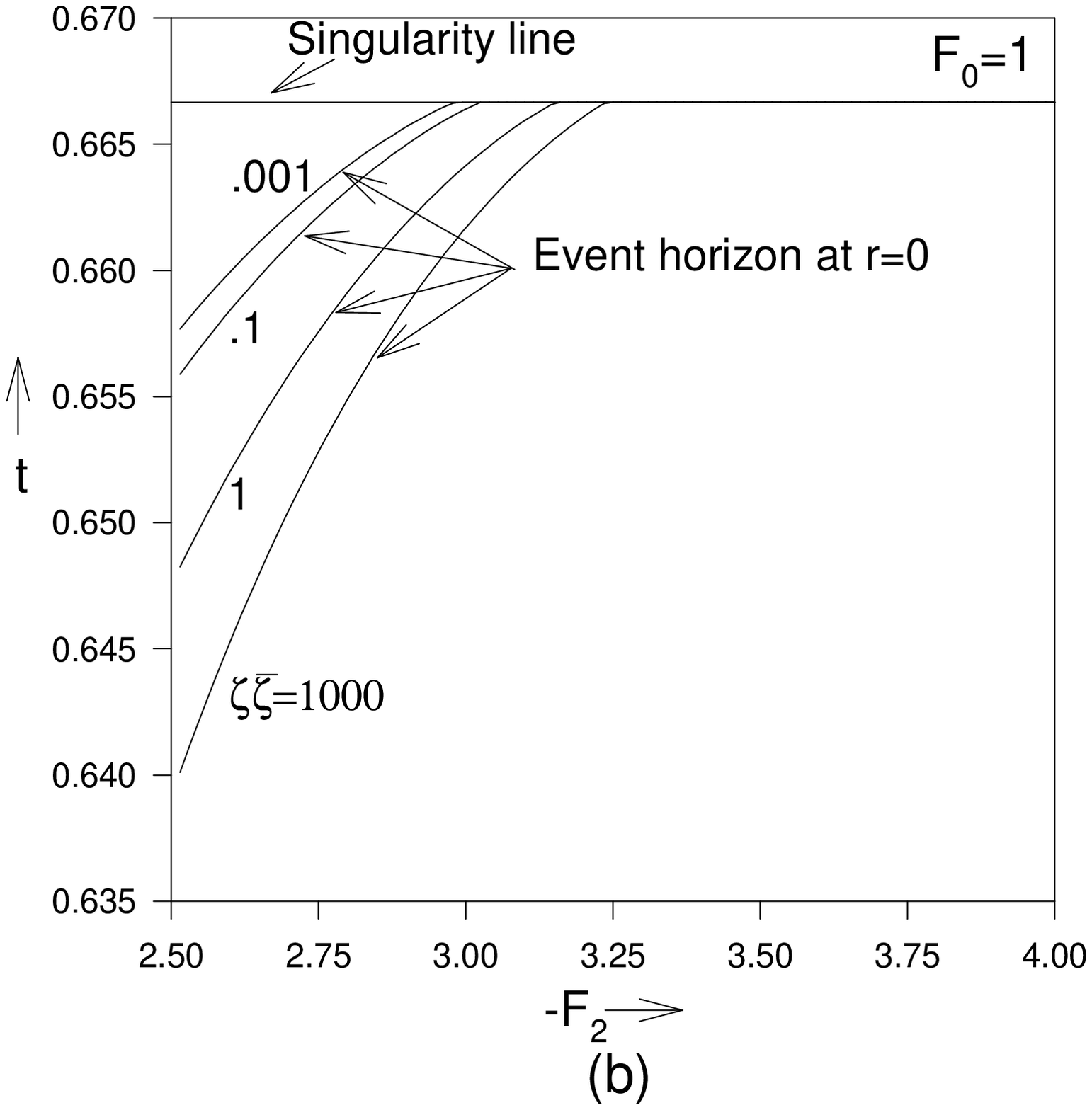}
%%{vasz2.ps}
}
\end{figure}
\begin{figure}[p]  
\parbox[b]{7.88cm}
{
\epsfxsize=7.85cm
\epsfbox{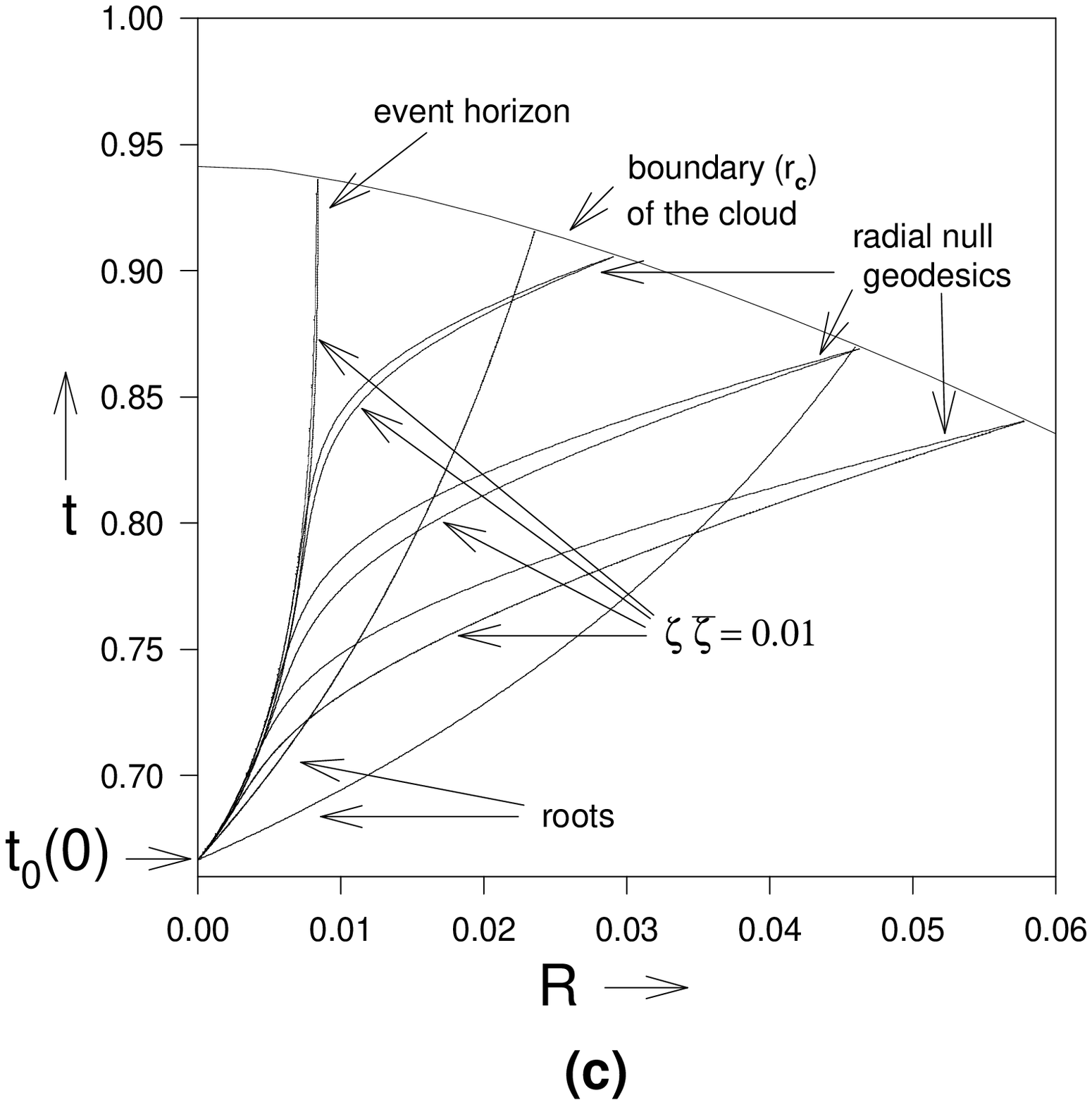}
}
\caption{
The graph showing the transition from locally naked to globally naked 
singularity in the marginally bound (f=0) Szekeres quasi-spherical models 
for different values of $\zeta \bar\zeta$. Singularity is located at 
$t=t_0(0)=2/3$. (a) $F_1\ne 0$, a = 1.0, boundary of the cloud is at $r_c 
=-3F_0/4F_1$. (b) $F_{2}\ne 0$, a = 2.0, boundary of the cloud is at 
$r_c=(-3F_0/5F_2)^{1/2}$. (c) The globally naked singularity in the case 
$F_3\ne 0$, a = 3.9.}
\end{figure}

\subsubsection {$F_2 \ne 0$}

The singularity is always locally naked and there is a family of 
RNGs with tangent as the apparent horizon near the singularity.
The general behavior of RNGs is similar to that of $F_1$ case
above. If we take the
boundary of the cloud at $r_c=\sqrt{-3F_0/5F_2}$, then in the range
$-3.25 <F_2 <-3.00 $ the singularity is globally naked depending on the value
of $\zeta \bar{\zeta}$ (Fig. 4(b)). That is, there is a mode of directional
global visibility for the singularity.
For $F_2< -3.25$ the singularity is globally visible independent of the 
direction,
and for $F_2>-3.00 $ the singularity is not globally 
visible for any $\zeta \bar{\zeta}$ values.

\subsubsection {$F_3 \ne 0$}  

In this case $\alpha= 3$, and we have two real positive roots for 
the equation $V(X)=0$ if $\xi={F_3}/{F_0}^{5/2} < -25.99038=\xi_{crit}$
and the singularity is locally naked. We see in this case that if the 
singularity is locally naked then it is always globally naked. 
We can also see numerically that there is a family of geodesics meeting the 
singularity with $X_0$ as the tangent (Fig. 4(c)).
But still the equation of trajectories has some dependence on the angle.

\section {Conclusions }

We have examined here the global visibility and structure of the
singularity forming in the gravitational collapse of a dust cloud, in the
spherically symmetric TBL case, and also for the aspherical Szekeres models.
An interesting feature that emerges, for the marginally bound case, is when
we depart from the homogeneous black hole case by introducing a small 
amount of density perturbation which makes the density distribution
inhomogeneous, even though the singularity becomes locally naked the global
asymptotic predictability is still 
preserved. Such a singularity is not globally
naked as our numerical results show, till the density perturbation
increases and 
crosses a certain critical value, and thus for small enough density
perturbations to the original homogeneous density profile corresponding
to the black hole case, the weak cosmic censorship is preserved. This,
in a way establishes the stability of the Oppenheimer-Snyder black hole,
when the mode of perturbation is that of the density fluctuations in the 
initial data. It may be noted that this is a different sort of stability
as opposed to the stability of the Schwarzschild black hole often referred
to in the literature. What we have in mind here is the stability of the
formation of black holes in gravitational collapse, involving the
inhomogeneities of matter distribution, which remains invariant under  
small enough density perturbations in the initial density profile from
which the collapse evolves.

It remains an interesting question to be examined further as to whether
this result will be still intact when we include the nonmarginally bound
case, involving the nonzero values of the energy function $f$. 

Our analysis also confirms the earlier results \cite{JosKro},
deducing in a more specific manner that under physically reasonable initial 
conditions naked singularities do develop in the Szekeres collapse 
space-times, which are not spherically symmetric, and admit no Killing 
vectors. This indicates the possibility that the earlier results for the 
final fate of 
spherical gravitational collapse might be valid when suitable generalizations 
to nonspherical spacetimes are made.

As we have shown here, for the marginally bound TBL and Szekeres models,
the global visibility of the singularity is related to the initial data. We 
classified the range of initial data for expandable mass functions 
$F(r)$. We see that if the density decreases ``fast enough", that is,
when there is sufficient inhomogeneity, then the 
singularity is globally naked. In the Szekeres models, for some critical 
range of initial data we see that the singularity is globally naked only
for a certain range of $\zeta\bar{\zeta}$. Hence the singularity will be 
globally visible only in some directions. But there is no such directional 
dependence as far as the local visibility of the singularity is concerned. 
Thus our results point out and analyze the global nakedness of the singularity 
for the collapse of a dust cloud which is either spherical or quasi-spherical,
with a regular initial data defined on a regular initial slice.

\end{document}